\documentclass[a4paper,fleqn]{article}
\usepackage{amsmath}
\usepackage[dvips]{graphicx}
\begin{document}

\title{\textbf{A new Painlev{\'e}-integrable equation \\
possessing KdV-type solitons}}

\author{\textsc{Sergei Sakovich}\bigskip \\
\small Institute of Physics, National Academy of Sciences of Belarus \\
\small sergsako@gmail.com}

\date{}

\maketitle

\begin{abstract}
A new three-dimensional second-order nonlinear wave equation is introduced which passes the Painlev{\'e} test for integrability and possesses KdV-type multisoliton solutions. Lax integrability of this equation remains unknown.
\end{abstract}

\section{Introduction}

In this paper, we study the following three-dimensional second-order nonlinear wave equation:
\begin{equation}
u_{xz} + u_{yy} + 2 u u_{xy} + 6 u^2 u_{xx} + 2 u_{xx}^2 = 0 . \label{e1}
\end{equation}
This equation appeared recently, as an interesting special case, in our work on the Painlev{\'e} analysis of a class of second-order equations quadratic in $u_{xx}$. The analysis of the whole class is still in progress. Therefore, in the present paper, we consider this new equation \eqref{e1} separately, as it is.

In Section~\ref{s2}, we show that the new equation \eqref{e1} passes the Painlev{\'e} test for integrability of nonlinear partial differential equations. In Section~\ref{s3}, we show that the nonlinear wave equation \eqref{e1} possesses multisoliton solutions which satisfy, simultaneously, the Korteweg--de~Vries (KdV) equation and the fifth-order KdV equation. In Section~\ref{s4}, we discuss the obtained results and some possible ways to study this new equation further.

\section{Painlev{\'e} analysis} \label{s2}

Let us show that the nonlinear wave equation \eqref{e1} passes the Painlev{\'e} test for integrability, in the formulation for nonlinear partial differential equations \cite{WTC,H}. 

The partial differential equation \eqref{e1} has order two and dimension three, therefore its general solution must contain two arbitrary functions of two variables. To start the Painlev{\'e} test, we use the expansion
\begin{equation}
u = u_0 (y,z) \phi^s + \dotsb + u_r (y,z) \phi^{r+s} + \dotsb \label{e2}
\end{equation}
for solutions $u(x,y,z)$ near the singularity manifold $\phi(x,y,z) = 0$ with $\phi_x = 1$, $\phi = x + f(y,z)$. Then we find from \eqref{e1} that the dominant singular behavior of solutions is determined by
\begin{equation}
s = -2 , \qquad u_0 = -2 , \label{e3}
\end{equation}
and that the values of resonances $r$, where arbitrary functions can enter the expansion \eqref{e2}, are
\begin{equation}
r = -1 , 6 , \label{e4}
\end{equation}
which correspond to the arbitrariness of $f(y,z)$ and $u_6 (y,z)$. Since the leading exponent $s$ \eqref{e3} and the resonances $r$ \eqref{e4} are integer, we can use the Laurent-type expansion
\begin{equation}
u = \sum_{i=0}^{\infty} u_i (y,z) \phi^{i-2}  \label{e5}
\end{equation}
near $\phi(x,y,z) = 0$, with $\phi = x + f(y,z)$.

Now we substitute the expansion \eqref{e5} to \eqref{e1}, collect terms with $\phi^{j-8}$, successively for $j = 0, 1, 2, \dotsc , 6$, and obtain in this way the following. For $j = 0$, of course, we get $u_0 = -2$, as in \eqref{e3}. For $j = 1, 2, 3, 4, 5$, we get, respectively,
\begin{gather}
u_1 = 0 , \label{e6} \\
u_2 = - \frac{1}{6} f_y , \label{e7} \\
u_3 = 0 , \label{e8} \\
u_4 = \frac{1}{20} f_z + \frac{1}{24} f_y^2 , \label{e9} \\
u_5 = - \frac{1}{36} f_{yy} . \label{e10}
\end{gather}
Finally, for $j = 6$, we get identically zero. All the terms with $\phi^{-2}$ are canceled out in \eqref{e1}, provided that the coefficients $u_0 , u_1 , \dotsc , u_5$ of \eqref{e5} are determined by \eqref{e3} and \eqref{e6}--\eqref{e10}. The coefficient $u_6 (y,z)$ and the function $f(y,z)$ of the singularity manifold remain arbitrary. In other words, the recursion relations for the coefficients $u_i$ of the Laurent-type expansion \eqref{e5} are consistent at the resonance $r = 6$, and there is no need to modify the expansion by additional non-dominant logarithmic terms. The nonlinear wave equation \eqref{e1} has passed the Painlev{\'e} test for integrability.

The reliability of the Painlev\'{e} test has been verified empirically in many works concerning the integrability of large classes of nonlinear equations, such as the fifth-order KdV-type equations \cite{HO}, the bilinear equations \cite{GRH}, the coupled KdV equations \cite{K,S1} (extra details in \cite{S2,KKS1}), the symmetrically coupled higher-order nonlinear Schr\"{o}dinger equations \cite{ST}, the generalized Ito equations \cite{KKS2}, the sixth-order bidirectional wave equations \cite{KKSST}, and the seventh-order KdV-type equations \cite{X}. Therefore we believe that the nonlinear wave equation \eqref{e1} must be integrable in the Lax sense. However, we have not found any Lax representation for this new equation as yet.

\section{Soliton solutions} \label{s3}

Let us show that the nonlinear wave equation \eqref{e1} possesses multisoliton solutions which satisfy, simultaneously, the KdV equation and the fifth-order KdV equation, and explain why this occur.

It is easy to find the one-soliton solution of \eqref{e1} directly. We look for a solution of \eqref{e1} in the form
\begin{equation}
u(x,y,z) = v(w) , \qquad w = x + a y + b z ,  \label{e11}
\end{equation}
where $a$ and $b$ are constants, and require that
\begin{equation}
v \to 0 \quad \text{for} \quad  w \to \pm \infty .  \label{e12}
\end{equation}
In this way, we get the ordinary differential equation
\begin{equation}
v'' + 3 v^2 + a v = 0 ,  \label{e13}
\end{equation}
where the prime denotes the derivative, and the relation
\begin{equation}
b = - a^2  \label{e14}
\end{equation}
which follows from \eqref{e12}. Then we multiply \eqref{e13} by $2 v'$, integrate with respect to $w$, set the constant of integration to be zero due to \eqref{e12}, and get
\begin{equation}
{v'}^2 + 2 v^3 + a v^2 = 0 .  \label{e15}
\end{equation}
Finally, we notice that $a$ must be negative due to \eqref{e12} and \eqref{e15}, set
\begin{equation}
a = - p^2 ,  \label{e16}
\end{equation}
solve the first-order equation \eqref{e15} (a constant of integration $c$ appears), and obtain the following one-soliton solution of the new equation \eqref{e1}:
\begin{equation}
u = \frac{1}{2} p^2 \operatorname{sech}^2 \left( \frac{1}{2} \left( p x - p^3 y - p^5 z - c \right)  \right) ,  \label{e17}
\end{equation}
where $p$ and $c$ are arbitrary constants.

We can immediately recognize the obtained expression \eqref{e17} as the well-known one-soliton solution of the KdV equation
\begin{equation}
u_y + u_{xxx} + 6 u u_x = 0 ,  \label{e18}
\end{equation}
for which the variable $z$ serves as a constant. In their singularity structure, solutions of the KdV equation \eqref{e18} are very similar to solutions of the new equation \eqref{e1}. In the KdV case, we have the same second-order pole at $\phi = 0$ and even the same expressions for the coefficients $u_0$, $u_1$, $u_2$, $u_3$ and $u_5$ as given by \eqref{e3}, \eqref{e6}, \eqref{e7}, \eqref{e8} and \eqref{e10}, while $u_4$ remains arbitrary because the resonances are $r = -1, 4, 6$ for the KdV equation. In both cases, \eqref{e1} and \eqref{e18}, the singular expansions of solutions start as
\begin{equation}
u = 2 ( \log \phi )_{xx} + \dotsb .  \label{e19}
\end{equation}
In the KdV case, this expression \eqref{e19} is sometimes considered as an indication to use the substitution
\begin{equation}
u = 2 ( \log \tau )_{xx}  \label{e20}
\end{equation}
which brings the KdV equation into its bilinear form \cite{H,GRTW}. The nonlinear wave equation \eqref{e1}, however, is not related by this substitution \eqref{e20} to any bilinear equation, as far as we see. Nevertheless, the one-soliton solution \eqref{e17} of \eqref{e1} can be expressed as \eqref{e20} with $\tau$ given by
\begin{equation}
\tau = 1 + \exp \left( p x - p^3 y - p^5 z - c \right) . \label{e21}
\end{equation}

For the above reason, let us use the substitution \eqref{e20} to find a two-soliton solution of the new equation \eqref{e1}. We take $\tau$ in the form
\begin{equation}
\tau = 1 + \exp \eta_1 + \exp \eta_2 + q \exp ( \eta_1 + \eta_2 ) , \label{e22}
\end{equation}
where
\begin{equation}
\eta_i = p_i x - p_i^3 y - p_i^5 z - c_i , \label{e23}
\end{equation}
and $q , p_i , c_i$ are constants, $i=1,2$. Then we find that the function $u(x,y,z)$ determined by \eqref{e20}, \eqref{e22} and \eqref{e23} is a solution of the new equation \eqref{e1} if the constant $q$ is determined by
\begin{equation}
q = \frac{( p_1 - p_2 )^2}{( p_1 + p_2 )^2} , \label{e24}
\end{equation}
for any constants $p_i$ and $c_i$, $i=1,2$. Surprisingly, the obtained two-soliton solution of \eqref{e1} is also the two-soliton solution of the KdV equation \eqref{e18}, for which the variable $z$ serves as a constant.

\begin{figure}
\includegraphics[width=12cm]{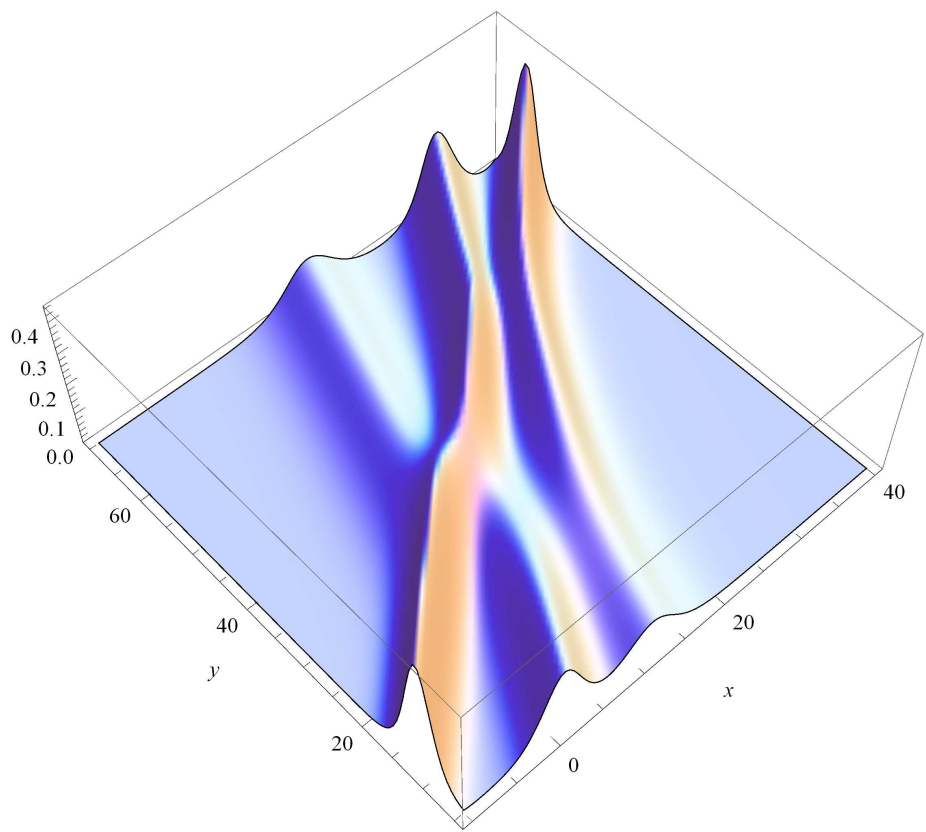}
\caption{A typical three-soliton solution of \eqref{e1}, at $z = -30$. \label{f1}}
\end{figure}

Now we are ready to suppose that the multi-soliton solutions of the new equation \eqref{e1}, if they exist, have the same form as the multi-soliton solutions of the KdV equation \eqref{e18}, except for the dependence on $z$ which should be introduced in the same way as in \eqref{e23}. We use the Hirota's representation for the $N$-soliton solution of the KdV equation \cite{Hir}, and look for the $N$-soliton solution $u(x,y,z)$ of \eqref{e1} in the form of \eqref{e20} with
\begin{equation}
\tau = \det M , \qquad M_{ij} = \delta_{ij} + \frac{2 ( p_i p_j )^{1/2}}{p_i + p_j} \exp \left( \frac{1}{2} ( \eta_i + \eta_j ) \right) , \label{e25}
\end{equation}
where $i, j = 1, 2, \dotsc , N$, expressions for $\eta_i$ are given by \eqref{e23}, and $p_i , c_i$ are arbitrary constants. The cases of $N=1,2$ have already appeared above as the one- and two-soliton solutions of \eqref{e1}. For $N=3,4,5,6$, we were able to verify by computer algebra means that the function $u(x,y,z)$ determined by \eqref{e20}, \eqref{e25} and \eqref{e23} does satisfy the nonlinear wave equation \eqref{e1}, for any values of the parameters $p_i$ and $c_i$ involved. A typical three-soliton solution of \eqref{e1} is shown in Figure~\ref{f1} for a fixed value of $z$, $z = -30$, with $p_1 = 0.5$, $p_2 = 0.7$, $p_3 = 0.9$ and $c_1 = c_2 = c_3 = 0$.

Let us show why the obtained multi-soliton solutions of the new equation \eqref{e1} are so closely related to the multi-soliton solutions of the KdV equation \eqref{e18}, although the new equation is (very likely) not related to the spectral problem of the KdV equation. We can rewrite the nonlinear wave equation \eqref{e1} as
\begin{gather}
\left( \partial_y - \partial_x^3 - 4 u \partial_x - 6 u_x \right) ( u_y + u_{xxx} + 6 u u_x ) \notag \\
\qquad \quad + \partial_x \left( u_z + u_{xxxxx} + 10 u u_{xxx} + 20 u_x u_{xx} + 30 u^2 u_x \right) = 0 . \label{e26}
\end{gather}
Now we see that any function $u(x,y,z)$ which simultaneously satisfies the KdV equation \eqref{e18} and the fifth-order KdV equation
\begin{equation}
u_z + u_{xxxxx} + 10 u u_{xxx} + 20 u_x u_{xx} + 30 u^2 u_x = 0 \label{e27}
\end{equation}
is a solution of the new equation \eqref{e1} as well. In other words, the over-determined system of equations \eqref{e18} and \eqref{e27} is a reduction of the new equation \eqref{e1}. The KdV equation \eqref{e18} and the fifth-order KdV equation \eqref{e27} are two members of the hierarchy associated with the spectral problem $\psi_{xx} + ( u + \lambda ) \psi = 0$, the $x$-dependence of the multi-soliton solutions given by \eqref{e20}, \eqref{e25} and \eqref{e23} is determined by the reflectionless potentials $-u$ of this spectral problem \cite{Hir,KM}, while the dependence on $y$ and $z$ in \eqref{e23} is determined by the dispersion relations of equations \eqref{e18} and \eqref{e27} via the ``time'' evolution of scattering data, the ``time'' being $y$ for \eqref{e18} and $z$ for \eqref{e27}. Consequently, the fact that the nonlinear wave equation \eqref{e1} possesses multi-soliton solutions tells nothing about integrability of this equation, it only tells that the equation possesses an integrable reduction.

\section{Discussion} \label{s4}

In this paper, we have introduced the new nonlinear wave equation \eqref{e1} which passes the Painlev{\'e} test for integrability and possesses KdV-type multisoliton solutions. The problem to find a Lax representation for the new equation still remains unsolved. Also, it would be interesting to study the new equation with respect to its symmetries, conservation laws, Hamiltonian structures, bilinear or multilinear representations, rational and periodic solutions, etc.

Of course, we tried to find a Lax pair for the new equation \eqref{e1}. One of the methods we used was the so-called truncation technique \cite{WTC,H}. Surprisingly, truncation of the singular expansion is compatible for the new equation \eqref{e1} (we omit complicated details) but produces only the well-known Painlev{\'e}--B{\"a}cklund transformations and Lax pairs for the KdV equation \eqref{e18} and the fifth-order KdV equation \eqref{e27}, not the desired results for the new equation \eqref{e1} itself. Such a phenomenon, when the truncation technique discovers an integrable reduction, can be seen in \cite{KS}, but the equation studied there is not Painlev{\'e}-integrable.

The over-determined system of equations \eqref{e18} and \eqref{e27} is a reduction of the new equation \eqref{e1}, in the sense that, due to the representation \eqref{e26}, any solution of the system of \eqref{e18} and \eqref{e27} is also a solution of \eqref{e1}. The opposite is not true. For example, the nonlinear equation \eqref{e1} admits the evident linear reduction $u_{xx} = u_{xy} = u_{xz} + u_{yy} = 0$, which immediately gives us the following class of special solutions of \eqref{e1}: $u = \alpha (z) x - \frac{1}{2} \alpha ' (z) y^2 + \beta (z) y + \gamma (z)$, where $\alpha$, $\beta$ and $\gamma$ are arbitrary functions, and the prime denotes the derivative. Of the whole this class, only the solution $u = \text{constant}$ satisfies the system of \eqref{e18} and \eqref{e27}. It would be interesting to find how many arbitrary functions (and of how many variables) are contained in the general solution of the over-determined system of equations \eqref{e18} and \eqref{e27}. This can be done by methods of the formal theory of differential equations \cite{Sei}.

And the last point to discuss concerns the representation \eqref{e26} of the new equation \eqref{e1}. It would be interesting to study whether (and in which sense) the nonlinear wave equation \eqref{e1} is an integrable extension of the KdV equation, similarly to the KdV6 equation \cite{KKSST} and its generalizations \cite{W,GP}.

\end{document}